\documentclass[10pt]{article}
\usepackage{ametsoc2col}
\newcommand\op[1]{\mathop{\rm #1}\nolimits}

\newcommand{\myabstract}{A linearized energy-balance model for global temperature is formulated, featuring a scale-free long-range memory (LRM) response and stochastic forcing representing the influence on the ocean heat reservoir from atmospheric weather systems. The model is parametrized by an effective response strength, the stochastic forcing strength, and the memory exponent. The instrumental global surface temperature record and the deterministic component of the forcing are used to estimate these parameters by means of the maximum-likelihood method. The residual obtained by subtracting the deterministic solution from the observed record is analyzed as a noise process and shown to be consistent with a long-memory time-series model and inconsistent with a short-memory model. By decomposing the forcing record in contributions from solar, volcanic, and anthropogenic activity one can estimate the contribution of each to 20'th century global warming. The LRM model is applied with a reconstruction of the forcing for the last millennium to predict the large-scale features of northern hemisphere temperature reconstructions, and the analysis of the residual also clearly favors the LRM model on millennium time scale. The decomposition of the forcing shows that volcanic aerosols give a considerably greater contribution to the cooling during the Little Ice Age than the reduction in solar irradiance associated with the Maunder minimum in solar activity. The LRM model implies a transient climate response in agreement with IPCC AR4 projections, but the stronger response on longer time scales suggests to replace the notion of equilibrium climate sensitivity by a time-scale dependent sensitivity.}
\begin{document}

%
%%%%%%%%%%%%%%%%%%%%%%%%%%%%%%%%%%%%%%%%%%%%%%%%%%%%%%%%%%%%%%%%%%%%%
% TITLE
%
% Enter your TITLE here
%%%%%%%%%%%%%%%%%%%%%%%%%%%%%%%%%%%%%%%%%%%%%%%%%%%%%%%%%%%%%%%%%%%%%
\title{\textbf{\large{Long-memory effects in linear-response models of Earth's temperature and implications for future global warming}}}
%
% Author names, with corresponding author information. 
% [Update and move the \thanks{...} block as appropriate.]
%
\author{\textsc{Martin Rypdal}\thanks{\textit{Corresponding author address:} 
				Martin Rypdal, University of Troms{\o}, N-9037 Troms{\o}, Norway.
				\newline{E-mail: martin.rypdal@uit.no}}\quad\textsc{and Kristoffer Rypdal}\\
\textit{\footnotesize{Department of Mathematics and Statistics, University of Troms{\o}, Norway.}}
}
%
% Formatting done here...Authors should skip over this.  See above for abstract.
\ifthenelse{\boolean{dc}}
{
\twocolumn[
\begin{@twocolumnfalse}
\amstitle

% Start Abstract (Enter your Abstract above.  Do not enter any text here)
\begin{center}
\begin{minipage}{13.0cm}
\begin{abstract}
	\myabstract
	\newline
	\begin{center}
		\rule{38mm}{0.2mm}
			\end{center}
\end{abstract}
\end{minipage}
\end{center}
\end{@twocolumnfalse}
]
}
{
\amstitle
\begin{abstract}
\myabstract
\end{abstract}
\newpage
}
%%%%%%%%%%%%%%%%%%%%%%%%%%%%%%%%%%%%%%%%%%%%%%%%%%%%%%%%%%%%%%%%%%%%%
% MAIN BODY OF PAPER
%%%%%%%%%%%%%%%%%%%%%%%%%%%%%%%%%%%%%%%%%%%%%%%%%%%%%%%%%%%%%%%%%%%%%
\section{Introduction}
 When the climate system is subject to  radiative forcing the planet is brought out of radiative balance and  the thermal inertia of the planet makes the surface temperature lag behind the forcing. The  time constant $\tau$, which is  the time for relaxation to a new equilibrium after a sudden change in forcing,  has been considered to be an important parameter to determine. The equilibrium  climate sensitivity $S_{\op{eq}}$, the temperature raise per unit forcing after relaxation is complete, is another. In the industrialized epoch  a major  source for the present energy imbalance is the steady increase in anthropogenic forcing.  If the climate system can be modeled as a hierarchy of interacting subsystems  with increasing heat capacities and response times there will also be a hierarchy of climate sensitivities. One way of modeling this feature is to replace the standard exponentially decaying impulse-response function $G(t)\sim e^{-t/\tau}$  with one that is scale free, i.e., decaying like a power law; $G(t)\sim t^{\beta/2-1}$. For a climate system which is subject only to random forcing modeled as a white Gaussian noise, and if $0<\beta<1$, the resulting climate variable $T(t)$ is then a long-memory fractional Gaussian noise (fGn) with a power spectral density (PSD) of the form ${\cal P}(f)\sim f^{-\beta}$ \color{blue} \citep{Beran:1994uu,Embrechts:2002tl}\color{black}. The response to a step at time $t=0$ in the forcing is then $\int_0^t G(t)\, dt\sim t^{\beta/2}$. Hence,  $S_{\op{eq}}$ is infinite for such a perfectly  scale-free response function, since the response to an increase in the forcing will never saturate. This is of course unphysical,  but rather than invalidating the scale-free response model it suggests the introduction of a frequency-dependent climate sensitivity $S(f)$. Even in the exponential response model the amplitude response to an oscillation   vansihes for high frequencies, but  converges to $S_{\op{eq}}$ as $f\rightarrow 0$. In the scale-free response model $S(f)$ diverges in the low-frequency limit, and hence  there is of course a cut-off frequency $f_c=1/t_c$ corresponding to a time-scale $t_c$  after which the scale-free response is no longer valid. In this paper, however,  we present evidence  for power-law scaling with $\beta \approx 1$  in the global temperature response over time scales of many centuries \color{blue} \citep{Rybski:2006bj,Rypdal:2013cc}\color{black}, suggesting that a lower bound for $t_c$ exceeds century time scales. We shall demonstrate  that long-memory responses can explain important aspects of Northern hemisphere temperature variability over the last millennium and lead to new predictions of  how much more warming there will be  ``in the pipeline" in any given forcing scenario \color{blue} \citep{Hansen:2005he,Hansen:2011ub}\color{black}.

Previous work on long-range memory (LRM) in climate records all hypothesize that the signal is composed of an LRM noise superposed on a trend driven by external forcing, and hence the methods are designed to eliminate such trends. The degree of detrending, however, is a parameter subject to choice. Choosing it too low implies that the detrending is incomplete, leading to overestimation of the memory exponent. Choosing it too high implies elimination of some of the internal noise and  underestimation of the memory. Another source of error is that the concept of a slow trend does not always reflect the true nature of deterministic forced variability. Some components of the forcing may  be faster than important components of the internal variability, and hence precise  separation of internal from forced variability can only be done by using information about the deterministic component of the forcing record. Fortunately, such reconstructions of the forcing records exist and are used as input for historic runs  of climate models. 

We contend that correct estimation of the LRM-proper-ties of the internal climate variability can only be done by analyzing the residue obtained by  subtracting the forced deterministic component of the climate signal. We shall show  that  the climate response function  is all we need to predict both the deterministic component of the climate signal and the memory properties of the internal variability.

\section{Linear response models}
Linear response models of Earth's surface temperature have been considered by several authors, see e.g. \color{blue} \cite{Hansen:2011ub} \color{black}  and  \color{blue} \cite{Rypdal:2012iya}\color{black}. The physical backbone is the zero-dimensional, linearized energy-balance equation derived for instance in the appendix of \color{blue} \citep{Rypdal:2012iya}\color{black}. It has the form
\begin{equation}
\frac{dQ}{dt}+\frac{1}{S_{\op{eq}}}T=F, \label{eqA1}
\end{equation}
where $Q$ is the total energy content of the climate system. $F$ and $T$ are perturbations of radiative influx and surface temperature relative to a reference state in radiative equilibrium, i.e., a state where the radiative influx absorbed by the Earth surface balances the infrared radiation emitted to space from the top of the troposphere. For a given influx the equilibrium outflux is controlled by the Stefan-Boltzmann radiation law and complex feedback processes which determine the equilibrium climate sensitvity $S_{\op{eq}}$ (see e.g., Eqs. (A5-A7)  in \color{blue} \cite{Rypdal:2012iya}\color{black}). The true value of $S_{\op{eq}}$ is subject to considerable controversy due to insufficient knowledge of  some of  these feedbacks, and because they operate on wildly different time scales.

The exponential response model is obtained by introducing an effective heat capacity $C$ of the climate system such that $dQ=CdT$, and introducing the time constant $\tau=CS_{\op{eq}}$. Eq.~(\ref{eqA1}) then takes the form
\begin{equation}
{\cal L}\, T=F, \label{eqA2}
\end{equation}
where the linear operator ${\cal L}\equiv C^{-1}(d_t+\tau^{-1})$ has the Green's function $G(t)=C^{-1}\exp{(-t/\tau)}$. The scale-free response model corresponds to replacing $\cal L$ by a fractional derivative operator (see \color{blue}\cite{Rypdal:2012iya} \color{black} for details) which effectively corresponds to replacing the exponential Green's function with the power-law function $ G(t)=(t/\mu)^{\beta/2-1}\xi$, where $\mu$ is a scaling factor in the units of time characterizing the strength of the response and $\xi\equiv 1$ Km$^2/$J is a factor needed to give $G(t)$ the right physical dimension.

We shall define our equilibrium reference state such that $T$ is the temperature relative to the initial temperature $\hat{T}_0$ in the observed record, i.e., $T=\hat{T}-\hat{T}_0$, where the $\hat{}$ symbol means that temperature is measured relative to absolute zero (Kelvin). The observed record then has $T(0)=0$. This means that we define $F$ as the influx relative to the influx which balances the outflux at this initial temperature, and since the system is not necessarily in equilibrium at $t=0$, we generally have that $F(0)\neq0$. Since $F(0)$ is not known a priori it becomes a parameter to be estimated along with other model parameters.
According to these conventions the temperature evolution according to linear response theory becomes
\begin{equation}
\hat{T}= \hat{T}_0+ F(0) \int_0^tG(t-s)\, ds+\int_0^tG(t-s)\tilde{F}(s)\, ds, \label{eqA3}
\end{equation}
where we have introduced $\tilde{F}(t)\equiv F(t)-F(0)$.
 In principle, we could have chosen $F(0)=0$ and $T(0)\neq 0$, but then we would for the scale-free response have to face  an initial value problem for a fractional differential equation, which is not well posed.
For the exponential response model Eq.~(\ref{eqA3}) takes the form
\begin{equation}
\hat{T}= \hat{T}_0+S_{\op{eq}} F(0)(1-e^{-t/\tau})+ \frac{1}{C}\int_0^te^{-(t-s)/\tau}\tilde{F}(s)ds. \label{eqA4}
\end{equation}
Suppose the evolution of the forcing prior to the time $t=0$ has given rise to an energy imbalance expressed through a non-zero $F(0)$, and that the forcing remains constant at this level ($\tilde{F}(t)=0$) in the subsequent evolution. Then equation (\ref{eqA4}) shows an exponential relaxation to a new equilibrium state with temperature $\hat{T}=\hat{T}_0+S_{\op{eq}}F(0)$. For the scale-free response model the corresponding expression takes the form
\begin{equation}
\hat{T}= \hat{T}_0+\frac{2\xi\mu F(0)}{\beta }\left(\frac{t}{\mu}\right)^{\beta/2}. \label{eqA5}
\end{equation}
The unlimited growth of the temperature in response to the initial step in forcing appears to be unphysical. For instance, Eq.~(\ref{eqA1})  implies that $dQ/dt$ goes negative for $t>t_c$, where \begin{equation} 
t_c\sim \mu \left(\frac{S_{\op{eq}}\beta}{2\xi\mu}\right)^{2/\beta}. \label{eqA6}
\end{equation}
One  solution to this paradox could be that the power-law tail in the response exhibits an exponential cut-off for $t>t_c$. We shall see, however,  that for generally accepted values of $S_{\op{eq}}$ the cut-off time is not more than a hundred years,  while we find evidence for power-law scaling up to a millennium at least. Another solution to the the paradox could be that Eq.~(\ref{eqA1}) is too simplistic. The equilibrium climate sensitivity is traditionally  defined from the end state of long model integrations subject to various forcing scenarios. It is difficult to decide if the models have attained equilibrium at the end of the integration, and $S_{\op{eq}}$ will depend critically on conventions  that distinguish between forcing and feedbacks. It is also conceivable that $S_{\op{eq}}$ is path-dependent, i.e, dependent on the forcing history. This issue is discussed further in section~6, after we have  established the full implications of the LRM response model.
\section{Dynamic-stochastic models}
In  \color{blue}\cite{Rypdal:2012iya} \color{black} it was shown that the scale-free response function gives a somewhat better characterization of the observed record, but no systematic method was presented which would allow rejection of the exponential response hypothesis in favor of the scale-free response hypothesis. The clue to develop such a method is to address the apparently random fluctuations in the observed record that makes it deviate from the  solution to the response model under the prescribed forcing. The forcing given by  \color{blue}\cite{Hansen:2011ub} \color{black}  is a deterministic function and the corresponding response  should therefore be perceived as a deterministic solution. Even with a correct model of the response the deterministic solution will not be a perfect match to the observed record because the forcing should also have a stochastic component corresponding to the random forcing of the ocean-land  heat content from the atmospheric weather systems. A more complete (dynamic-stochastic) model can be constructed by adding a stochastic forcing such that Eq.~(\ref{eqA2}) is generalized to
\begin{equation}
{\cal L}\, dT(t)=dF_d(t)+\sigma dB(t). \label{eq7a}
\end{equation}
Here $F_d(t)$ is the deterministic component of the forcing and $B(t)$ is the Wiener process, sometimes called a Brownian motion. We have introduced an unknown parameter $\sigma$ denoting the standard deviation of the stochastic forcing. There are two major  advantages of introducing the stochastic forcing:

 (i) Since the observed record in this formulation should be perceived as one realization of a stochastic process produced by the dynamic-stochastic model the residual difference between this record and the deterministic solution should be perceived as a noise process  $\tilde{T}(t)$ given by the stochastic part of Eq.~(\ref{eq7a}), i.e.,
\begin{equation}
\tilde{T}(t) =\sigma \int_{0}^ t G(t-s) dB(s). \label{eq6}
\end{equation}
By using the exponential response model, Eq.~(\ref{eq6}) produces the Ornstein-Uhlenbeck stochastic process. On time scales less than $\tau$ this process has the non-stationary character of a Brownian motion and the PSD has the power-law form ${\cal P}(f)\sim f^{-2}$ for $f>\tau^{-1}$. On time scales greater than $\tau$ the process has the stationary character of a white noise and the PSD is flat for $f<\tau^{-1}$. Actually, the PSD has the form of a Lorentzian,  ${\cal P}(f)\sim [\tau^{-2}+(2\pi f)^2]^{-1}$. For a discrete-time process the direct analog to the Ornstein-Uhlenbeck process is the first-order autoregressive process AR(1). The scale-free response model, on the other hand, produces a   fractional Gaussian noise (fGn) for $-1<\beta<1$ and a fractional Brownian motion (fBm) for $1<\beta<3$. For these noises and motions the PSD for low frequencies has the power-law form ${\cal P}(f) \sim f^{-\beta}$. In principle, an estimator for the PSD (like the periodogram) applied to the observed  residual could be compared to the PSD  for the two response models to test the validity of the models against each other. In practice, other estimators in this paper will be used, but the idea is the same.

(ii) Formulating the problem as a parametric  stochastic model allows systematic estimation of the parameters $\{F(0),C,\sigma, \tau\}$ for the exponential model,  and $\{F(0),\mu,\sigma, \beta\}$ for the scale-free model. The method is based on maximum-likelihood estimation (MLE) which establishes the most likely parameter set that could produce the observed record from the prescribed forcing. The principles of the MLE employed here are explained in the appendix.

The significance of the LRM response can be appreciated by looking at equation (\ref{eqA2}) in the Fourier domain;
\begin {equation}
\tilde{T}(f)=\tilde{G}(f)\tilde{F}(f), \label {eq7}
\end{equation}
where $\tilde{T},\tilde{G},\tilde{F}$ are Fourier transforms of $T,G,F$, and $\tilde{G}(f)$ is the transfer function of the linear system. This relation naturally leads  us to the define the frequency-dependent sensitivity as
\begin {equation}
S(f)=|\tilde{G}(f)|=\frac{|\tilde{T}(f)|}{|\tilde{F}(f)|}. \label {eq8}
\end{equation}
For the exponential response model we find
\begin {equation}
S(f)=\frac{1}{C\sqrt{\tau^{-2}+(2\pi f)^2}}, \label {eq9}
\end{equation}
which in the limit $2\pi f\tau\ll1$ converges to the equilibrium sensitivity $S_{\op{eq}}=\tau/C$. For the LRM model we have
\begin {equation}
S(f)=\frac{\xi \mu \Gamma (\beta/2)}{|2\pi \mu f|^{\beta/2}}, \label {eq10}
\end{equation}
where $\Gamma (x)$ is the Euler Gamma function. In Fig.~1 we show a plot of $S(f)$ for the values of  the model parameters estimated for the global temperature and forcing record in section~4. Note that the frequency-dependent sensitivities for the two models depart substantially from each other only for frequencies corresponding to time scales longer than a century. Hence it is on these slow time-scales that LRM really has serious impact on the climate dynamics. The dramatic consequences will be  apparent when we consider  time-scales of many centuries in section~5.

\begin{figure}[h]
\begin{center}
\includegraphics[width=6.5cm]{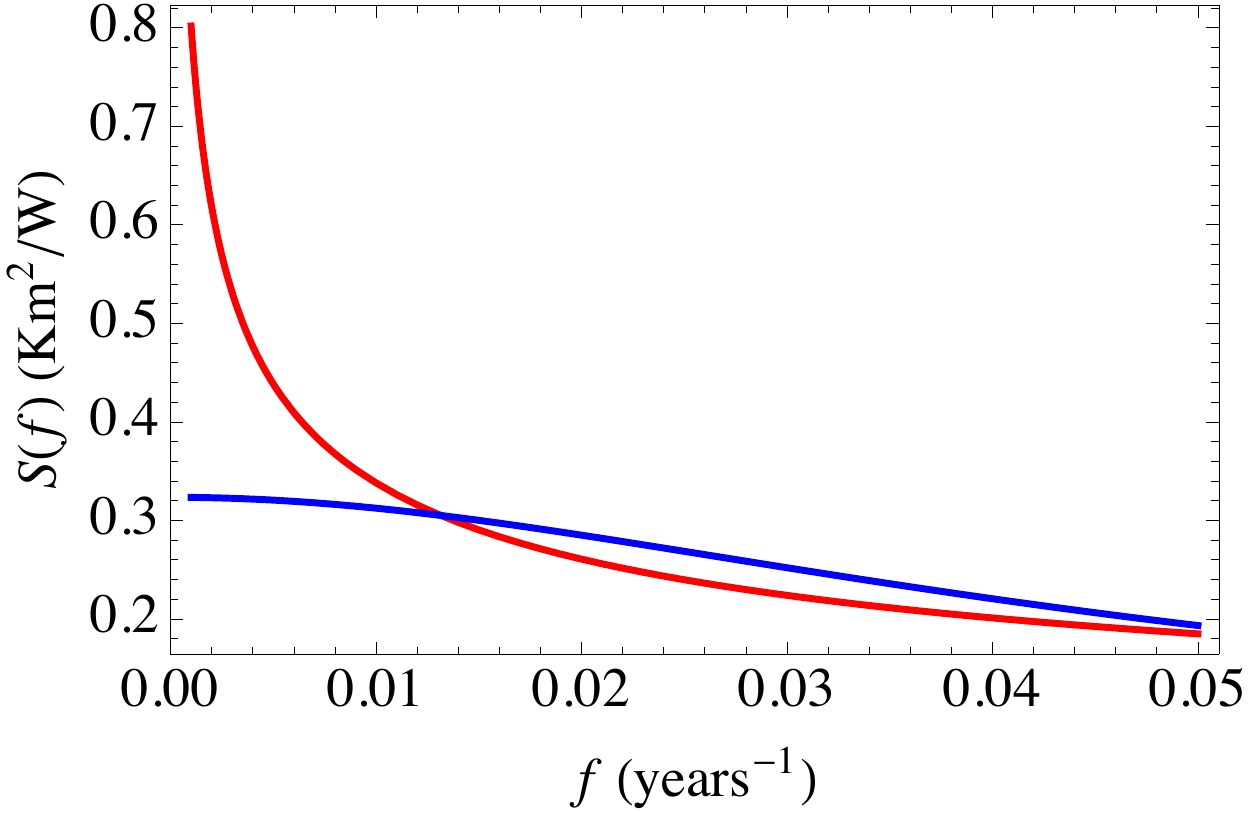}
\caption{Frequency-dependent sensitivity $S(f)$ for the exponential response model (blue) and the scale-free model (red) for model  parameters given in Tab.~1. } 
\end{center}
\end{figure}

In principle, the right-hand-side of Eq.~(\ref{eq8}) could be used to estimate $S(f)$ directly from Fourier transforming the temperature and forcing records, and then to compare with Eqs.~(\ref{eq9}) and (\ref{eq10}) to assess the validity of the two response models. The short length of the records, however, make the Fourier spectra very noisy, and the ratio between them even more so. Additional complications are that the spiky nature of  the forcing record to volcanic eruptions and the unknown amplitude of the stochastic forcing component. Hence, we have to resort to the model parameter estimation described  above, and to other estimators than the Fourier transform, to settle this issue.

\section{Parameter estimation from instrumental records}
The temperature data  sets analyzed in this section can be downloaded from the Hadley Center  Met Office web site. We  consider the global mean surface temperature (GMST) as presented by the HadCRUT3 monthly  mean or annual mean temperatures \color{blue} \citep{Brohan:2006wi}\color{black}. The forcing record is the one developed by \color{blue} \cite{Hansen:2005he}\color{black} and used by \color{blue}\cite{Hansen:2011ub} \color{black} and, and is shown in Fig.~2a. The forcings  decomposed into  volcanic, solar, and anthropogenic contributions  are shown in Figs. 2 b,c,d, respectively. The forcing records  go from 1880 till 2010 with annual resolution, so even if the temperature record goes further back in time and has monthly resolution, the maximum-likelihood estimation of model parameters only employs the 130 yr records with annual resolution. The analysis of the residual  noise signal, however, utilizes the monthly resolution to improve the statistics.

\begin{figure}[h]
\begin{center}
\includegraphics[width=9cm]{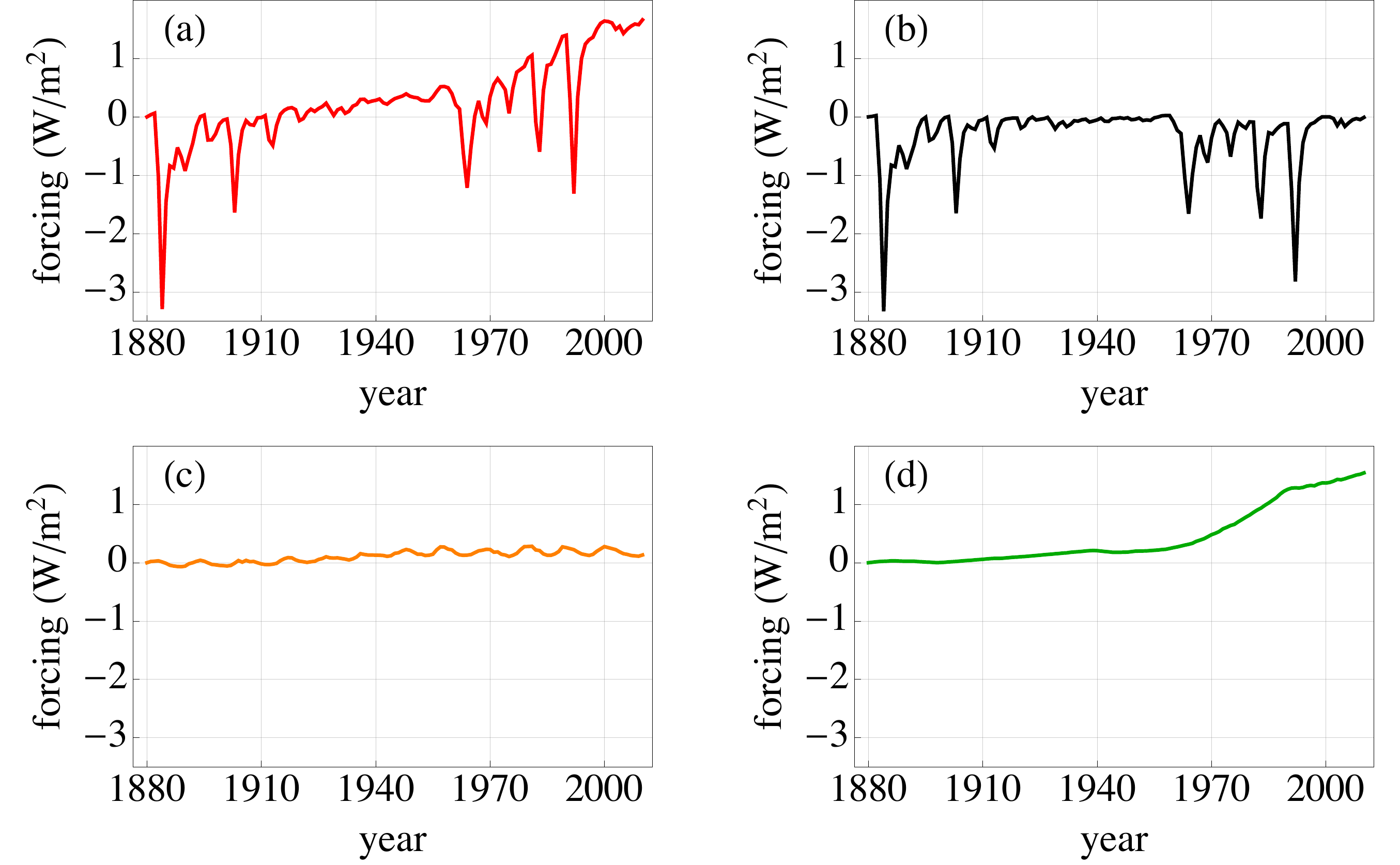}
\caption{(a) Total forcing 1880-2010 A.C. (b) Volcanic forcing. (c) Solar forcing. (d) Anthropogenic forcing.}
\end{center}
\end{figure}

\begin{table}[h]
\begin{tabular}{||l|l||} \hline 
{\small Exponential response model} & {\small Scale-free response model} \\ \hline \hline
{\small $\tau = 4.3 \,(\pm 0.7)$ years} & {\small $\beta = 0.75 \,(\pm 0.12)$} \\
{\small $C=4.2\, (\pm 0.2) \times 10^8$ J/m$^2$}  & {\small $\mu= 8.4 \,(\pm 2.5) \times 10^{-3}$ years} \\
{\small $F_0 = 0.19 \,(\pm 0.12)$ W/m$^2$} & {\small $F_0=0.19 \,(\pm 0.16)$ W/m$^2$} \\
{\small $\sigma_{\tilde{T}}=0.15\,(\pm 0.01)$ K} & {\small $\sigma_{\tilde{T}}=0.13 \,(\pm 0.02)$ K} \\ \hline \hline
\end{tabular} \caption{The ML estimates of parameters in the exponential response model and in the scale-free response model. The parameters are estimated from the HadCRUT3 annual temperature record. The parameters $\sigma_{\tilde{T}}$ are defined as the standard deviation of the stochastic components $\tilde{T}(t)$. The numbers in the brackets are the mean standard errors obtained from a Monte-Carlo study. }
\end{table}

The results of the MLE method  for the exponential and scale-free models are given in Tab.~1. The heat capacity $C=4.2\times 10^8$ J/m$^2$ estimated from the exponential model is very close to that of  a 100 m deep column of sea water, and the time constant $4.3$ yr is in the middle of the range (3-5 yr) observed  by \color{blue} \cite{Held:2013bi} \color{black} from instantaneous CO$_2$  experiments with the  CM2.1 model. What was also observed in those model runs was an additional slower response which showed that equilibrium was not attained after 100 yr of integration, indicating that the exponential model does not contain the whole story. In Fig.~3a we present the deterministic part of the solutions for both models along with the observed  GMST record. Although the the solution of the scale-free model seems to  yield  a  somewhat better representation of both the multidecadal variability and the response to volcanic eruptions, the difference between the deterministic  solutions of the  two models is not striking on these time scales. The reason for this can be understood from Fig.~1.  It is on time scales longer than a century that the difference will become apparent. For the stochastic part of the response, however,  the  two models can be tested against data on all observed time scales. Such a test is performed in Fig.~3b, where the residual noise (the observed GMST with the deterministic solution subtracted) has been analyzed by the wavelet-variance technique \color{blue} \citep{Flandrin:1992wt,Rypdal:2013cc}\color{black}. What is plotted here is the variance of the Mexican-hat wavelet coefficient of the residual noise versus wavelet scale. For an AR(1) process (stochastic solution of the exponential model) the the slope of this curve in a log-log plot is near 2 for time scales much less than $\tau$, and near 0 for time scales much greater than $\tau$, as shown by the blue dashed curve in the figure. For an fGn the slope of the curve is $\beta$, which has been estimated to be $\beta\approx 0.75$, as shown by the red dashed curve. The wavelet variance of the actual observed residuals with reference to the two models are shown as the blue crosses and the red circles in the figure, and shows that the residuals are inconsistent with an AR(1) process, but consistent with an fGn process.

\begin{figure}[h]
\begin{center}
\includegraphics[width=9cm]{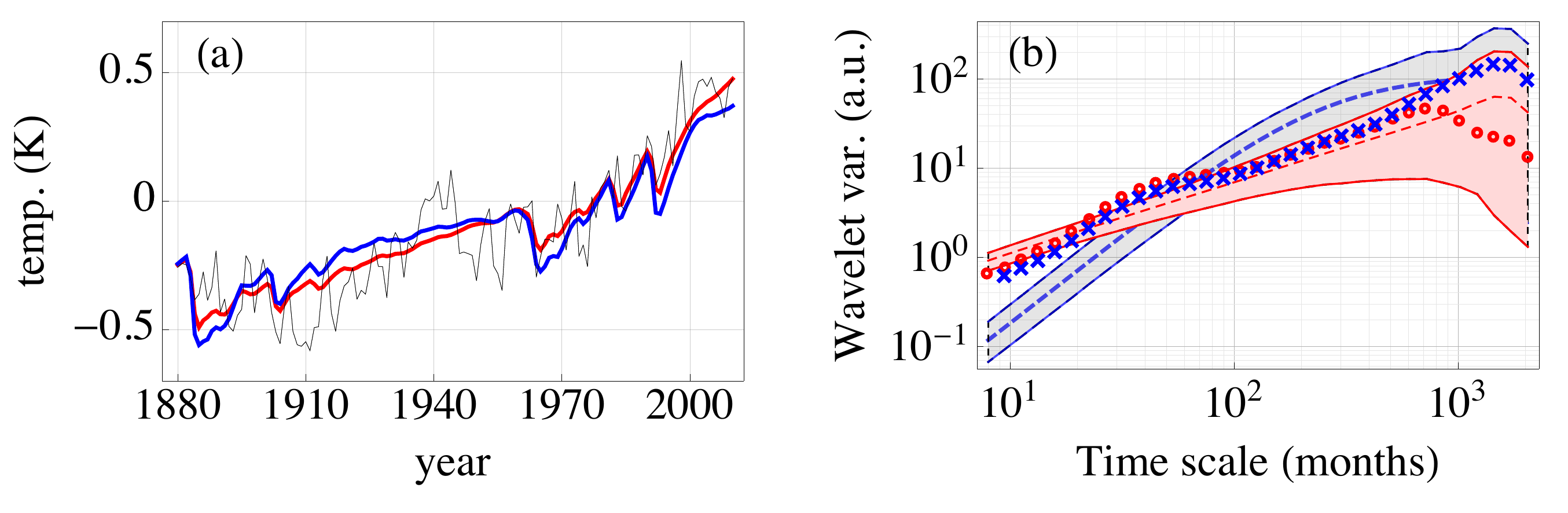}
\caption{(a) Deterministic part of the solution.  Blue: for the exponential response model. Red: for the scale-free response model. Black: the HadCRUT3 annual temperature record. (b) Blue crosses: Wavelet variance of monthly GMST record with deterministic solution of exponential response model subtracted.  Red circles: the same with deterministic solution of scale-free model subtracted. Blue dashed: Ensemble mean of wavelet variance of simulated AR(1) process with estimated parameters from the exponential response model. Shaded blue area marks  $2 \times$standard deviation of the distribution of wavelet variances over the ensemble. Red dashed and shaded area: the same for an fGn process with estimated parameters from the scale-free model. The wavelet used is the Mexican-hat wavelet, and the time scale on the horizontal axis is the wavelet-scale multiplied by 4, which corresponds closely to the oscillation period in the periodogram.}
\end{center}
\end{figure}

In Fig.~4 we demonstrate that the observed record falls within the uncertainty range of  the two dynamic-stochastic models. Here we have  generated an ensemble of solutions to the two models with the estimated parameters and plotted the $2\sigma$ range around the deterministic solutions. The results are shown as the two shaded areas in Figs.~ 4a and 4b, respectively.

\begin{figure}[h]
\begin{center}
\includegraphics[width=9cm]{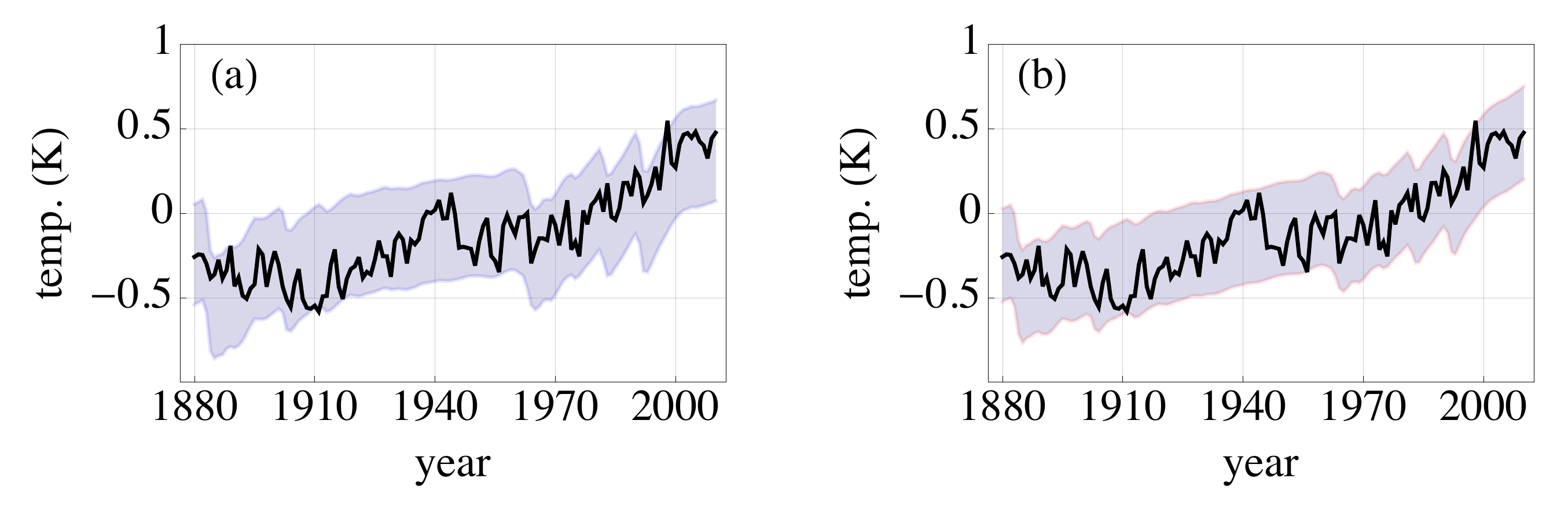}
\caption{ Black curves are the GMST record. Shaded  areas are the deterministic part of the solution $\pm 2\times$standard deviation of the stochastic part representing the range of solutions to the model. (a) for the exponential response model. (b) for the scale-free response model.}
\end{center}
\end{figure}
In Fig.~5 we plot the deterministic scale-free response to the total forcing along with the separate responses to the volcanic, solar, and anthropogenic forcing components. During the first half of the 20'th century solar and anthropogenic forcing contribute equally to the global warming trend. After 1950 there is a significant cooling trend due to volcanic aerosols, a weaker warming contribution from solar activity,  and a dominating anthropogenic warming.

\begin{figure}[h]
\begin{center}
\includegraphics[width=9cm]{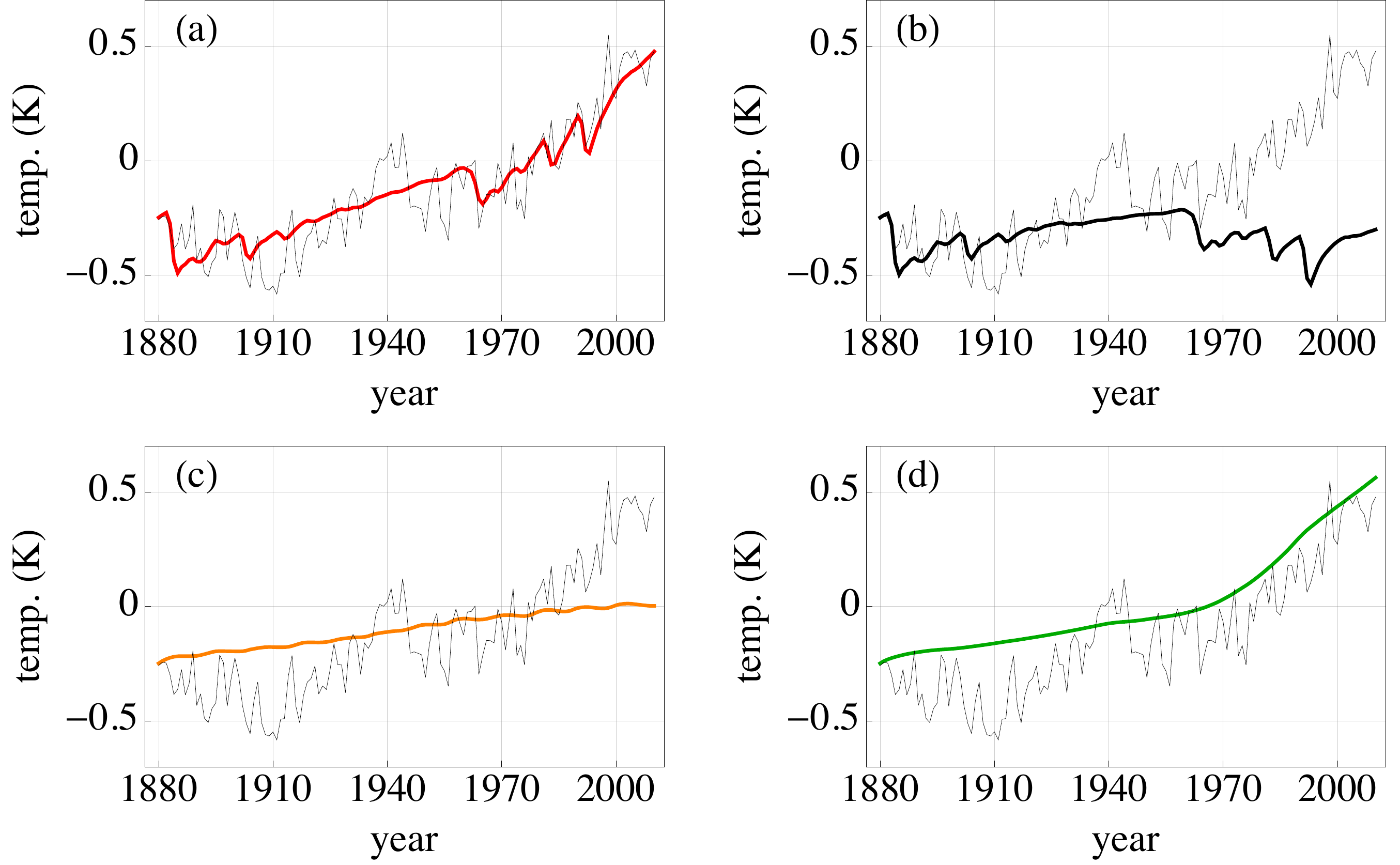}
\caption{Deterministic part of forced temperature change 1880-2010 A.C. according to the scale-free model.  (a) From  total forcing. (b) From volcanic forcing. (c) From solar forcing. (d) From anthropogenic forcing.}
\end{center}
\end{figure}
\section{Predicting reconstructed records }
The wavelet variance  plotted in Fig.~3b can only demonstrate that the residual is  scale free up to time scales less than  the length of the 130 yr record.  Verifying LRM on longer time scales requires longer records. This was done by \color{blue}\cite{Rybski:2006bj} \color{black} and \color{blue} \cite{Rypdal:2012iya} \color{black}  using detrending techniques like the wavelet variance applied directly to reconstructed temperature records over the last one or two millennia. Here we shall utilize a forcing record for the last millennium \color{blue} \citep{Crowley:2000du} \color{black} which is shown in Fig.~6, with its decomposition in volcanic, solar, and anthropogenic contributions.
\begin{figure}[h]
\begin{center}
\includegraphics[width=9cm]{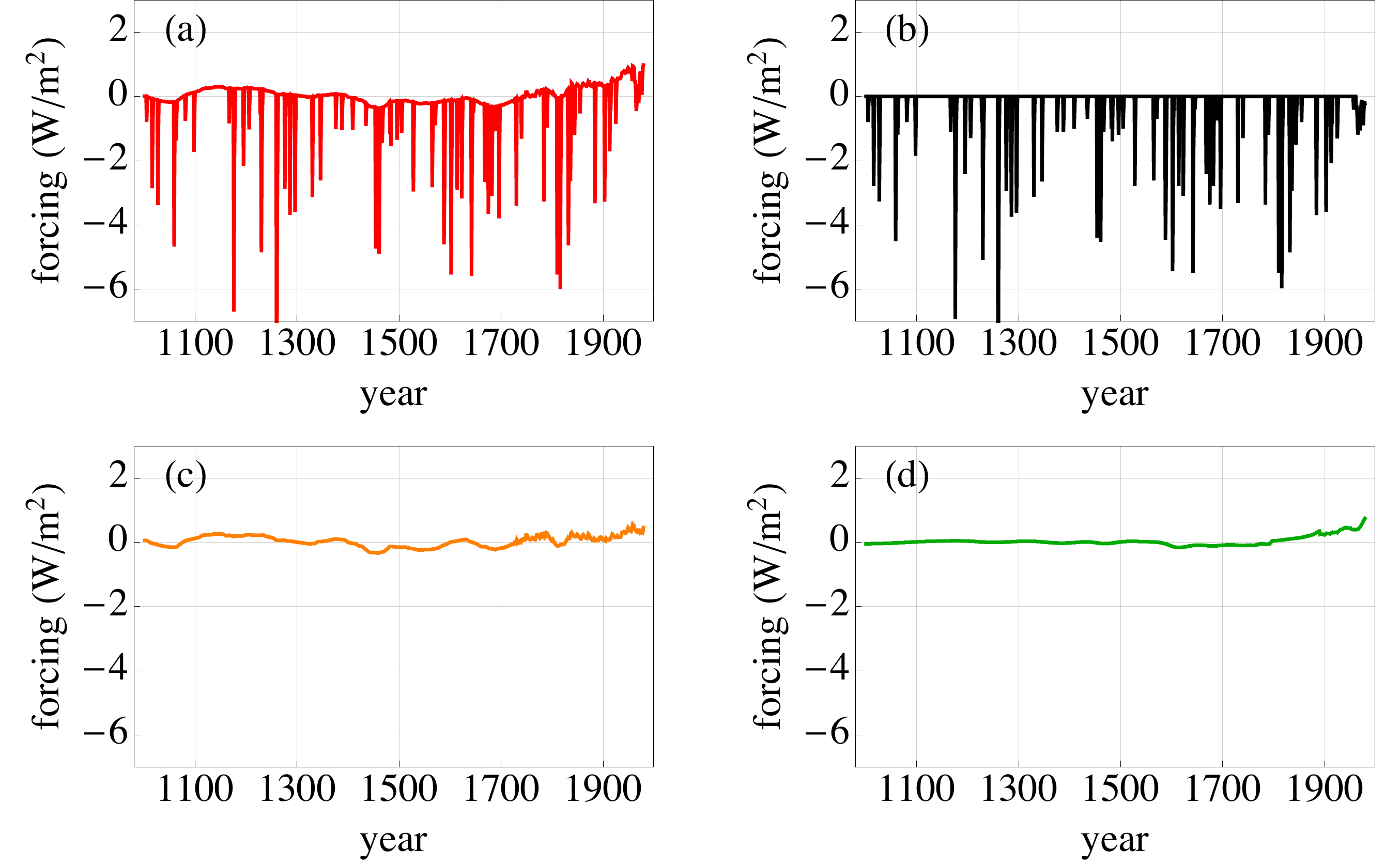}
\caption{(a) Total forcing 1000-1978 A.C. (b) Volcanic forcing. (c) Solar forcing. (d) Anthropogenic forcing.}
\end{center}
\end{figure}
Many temperature reconstructions for the Northern hemisphere exist for this time period (see \color{blue} \cite{Rybski:2006bj} \color{black} for a selection). We shall employ our dynamical-statistical models to the reconstruction by \color{blue} \cite{Moberg:2005wi}\color{black}, which  shows a marked temperature difference between the Medieval Warm Period (MWP) and the Little Ice Age (LIA). For the scale-free model the model parameters estimated from Crowley forcing and Moberg temperature are very  close to those estimated from the instrumental records, except for the initial forcing $F(0)$. The initial forcing measures how far the climate  system is from equilibrium at the beginning of the record,  and this will depend on at what time this beginning is chosen. Considering that the timing of volcanic events and the corresponding temperature responses probably are subject to substantial errors in these reconstructions, this might give rise to errors in the parameter estimates. For this reason we have also estimated $F(0)$ from Crowley forcing and Moberg temperature by  retaining the values of the other parameters estimated from the instrumental record and shown in Tab.~1. The resulting deterministic solutions for the two models are plotted in Fig.~7a, along with the Moberg record. Since only the departure from equilibrium forcing $F(0)$ are estimated from the reconstruction data, these solutions should be considered as ``predictions'' of the deterministic component of the forced evolution over the last millennium, based on parameters estimated from the modern instrumental records. The  exponential model  predicts too low temperature in the first half of the record and  too strong  short-term responses to volcanic eruptions. The scale-free model gives a remarkably good  reproduction of the large scale structure of the Moberg record and reasonable  short-term volcano responses. The wavelet variance of the residuals for the  two models are plotted in Fig.~7b, and again we observe that the results are consistent with a scale-free response over the millennium-long record and inconsistent with the exponential-response model.

\begin{figure}[h]
\begin{center}
\includegraphics[width=9cm]{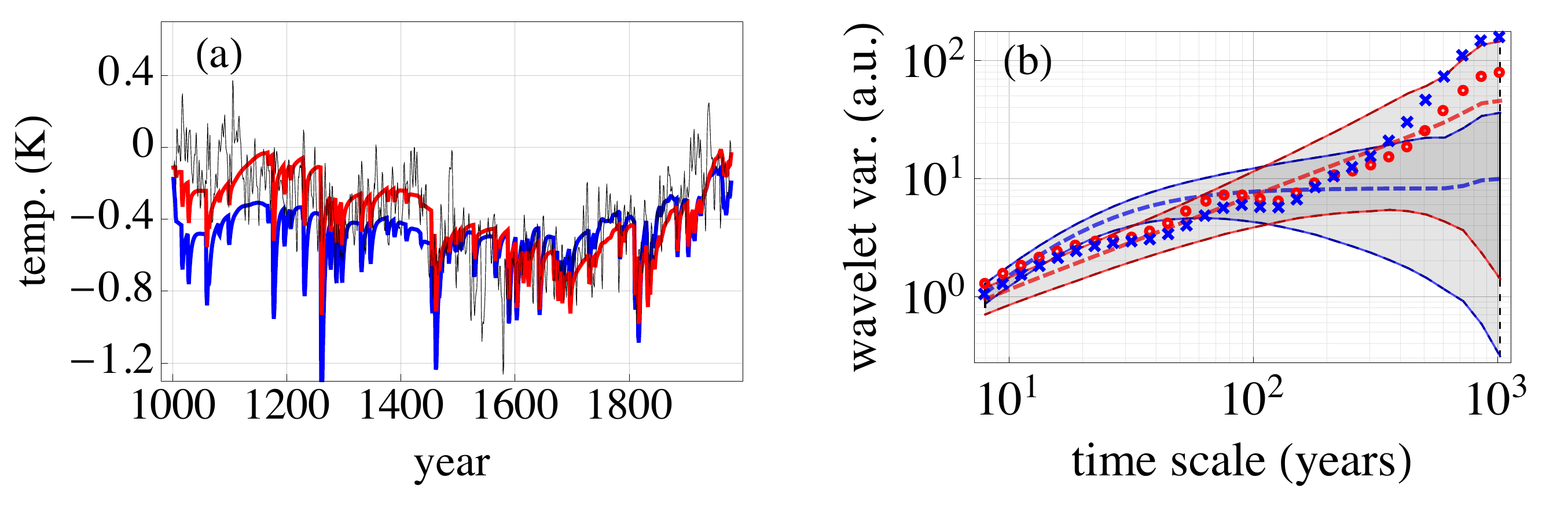}
\caption{(a) Deterministic part of the solution with Crowley forcing.  Blue: for the exponential response model. Red: for the scale-free response model. Black: the Moberg annual temperature reconstruction record. (b) Blue crosses: Wavelet variance of Moberg record with deterministic solution of exponential response model subtracted.  Red circles: the same with deterministic solution of scale-free model subtracted. Blue dashed: Ensemble mean of wavelet variance of simulated AR(1) process with estimated parameters from the exponential response model. Shaded blue area marks  $2 \times$standard deviation of the distribution of wavelet variances over the ensemble. Red dashed and shaded area: the same for an fGn process with estimated parameters from the scale-free model.}
\end{center}
\end{figure}

Fig.~8 shows the scale-free response to the total Crowley forcing, along with the responses to the volcanic, solar, and anthropogenic component. The most remarkable feature is that most of the cooling from the MWP to the LIA appears to be caused by volcanic cooling and not by the negative solar forcing associated with the Maunder minimum. On the other hand, the solar contribution to the warming from the LIA until mid 20'th century is comparable to the anthropogenic. After this time the warming is completely dominated by anthropogenic forcing, in agreement with what was shown in Fig.~5.

\begin{figure}[h]
\begin{center}
\includegraphics[width=9cm]{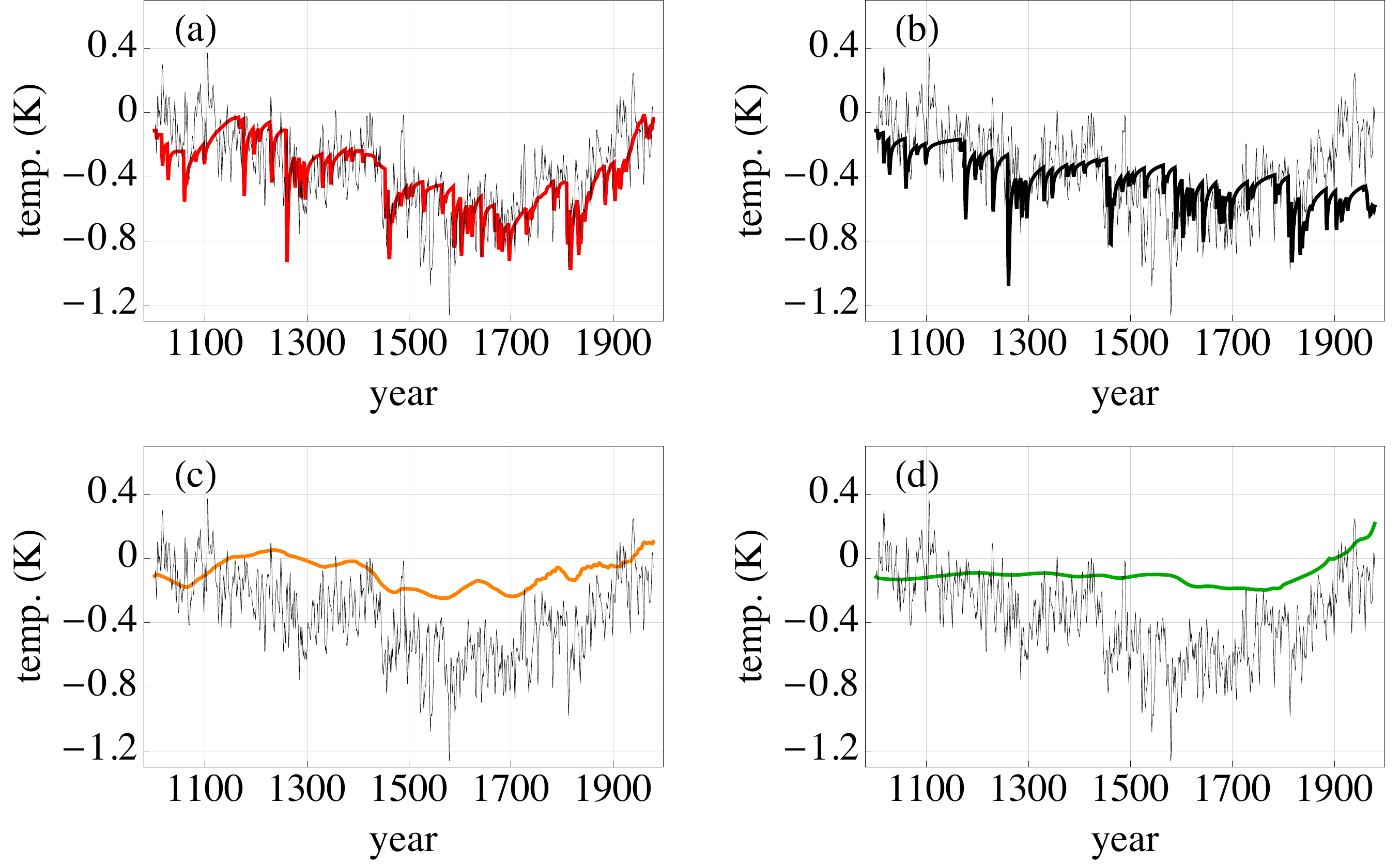}
\caption{Deterministic part of forced temperature change 1000-1978 A.D. according to the scale-free model with $F(0)$ estimated from the Moberg record and $\beta, \mu, \sigma$ from the instrumental record. Black curve is the Moberg record. (a) From  total forcing. (b) From volcanic forcing. (c) From solar forcing. (d) From anthropogenic forcing.}
\end{center}
\end{figure}

\section{Perspectives on climate sensitivity}
For predictions of future climate change on century time scales the equilibrium climate sensitivity may not be the most interesting concept. The frequency-dependent climate sensitivity $S(f)$ given by Eq.~(\ref{eq10}) is  more relevant. The transient climate response (TCR), defined as the temperature increase   $\Delta T_{\op{tr}}$ at the time of doubling of CO$_2$ concentration in a scenario where CO$_2$ concentration increases by 1\% per year from preindustrial levels, can also readily be computed from the response models. In Fig.~9a this forcing is shown as the dotted curve to the left (the forcing is logarithmic in the CO$_2$ concentration, so the curve is linear). The response curves to this forcing according to the two response models are  shown as the blue and red dotted curves to the left in panel (b). The end of these curves (at the time of CO$_2$ doubling after 70 yr) the temperatures represent the respective TCRs. They are both in the lower end of the range presented by the \color{blue}\cite{IntergovernmentPanelofClimateChange:wb}\color{black}. A more useful definition is to consider the response to the same CO$_2$ increase from the present  climate state  that is established from the historical forcing since preindustrial times. This response is what is shown as the blue and red full curves in Fig.~9b for the next 70 yr. For the scale-free model the temperature in year 2010 lags behind the forcing due to the memory effects, and the energy flux imbalance $dQ/dt$ established by the historical evolution at this time gives rise to a faster growth in the temperature during the next 70 yr, compared to the CO$_2$ doubling scenario starting in year 1880.  $\Delta T_{\op{tr}}$ (according to the modified definition)  is 1.3 K in the exponential response model, but  2.1 for  the scale-free model. The latter is very close to the median for the TCR given in \color{blue} \cite{IntergovernmentPanelofClimateChange:wb}\color{black}. Another illustration of the memory effect can be seen from the forcing scenario where the forcing is kept constant after 2010 as shown by the dashed line in Fig.~9a. The corresponding responses are given by the blue and red dashed curves in panel (b). The short time constant in the exponential model makes the temperature stabilize in equilibrium after a few years, while in the scale-free model the temperature keeps rising as $[2\mu^{1-\beta/2}F(2010)/\beta](t-2010)^{\beta/2}$ for $t>2010$ yr.

\begin{figure}[h]
\begin{center}
\includegraphics[width=9cm]{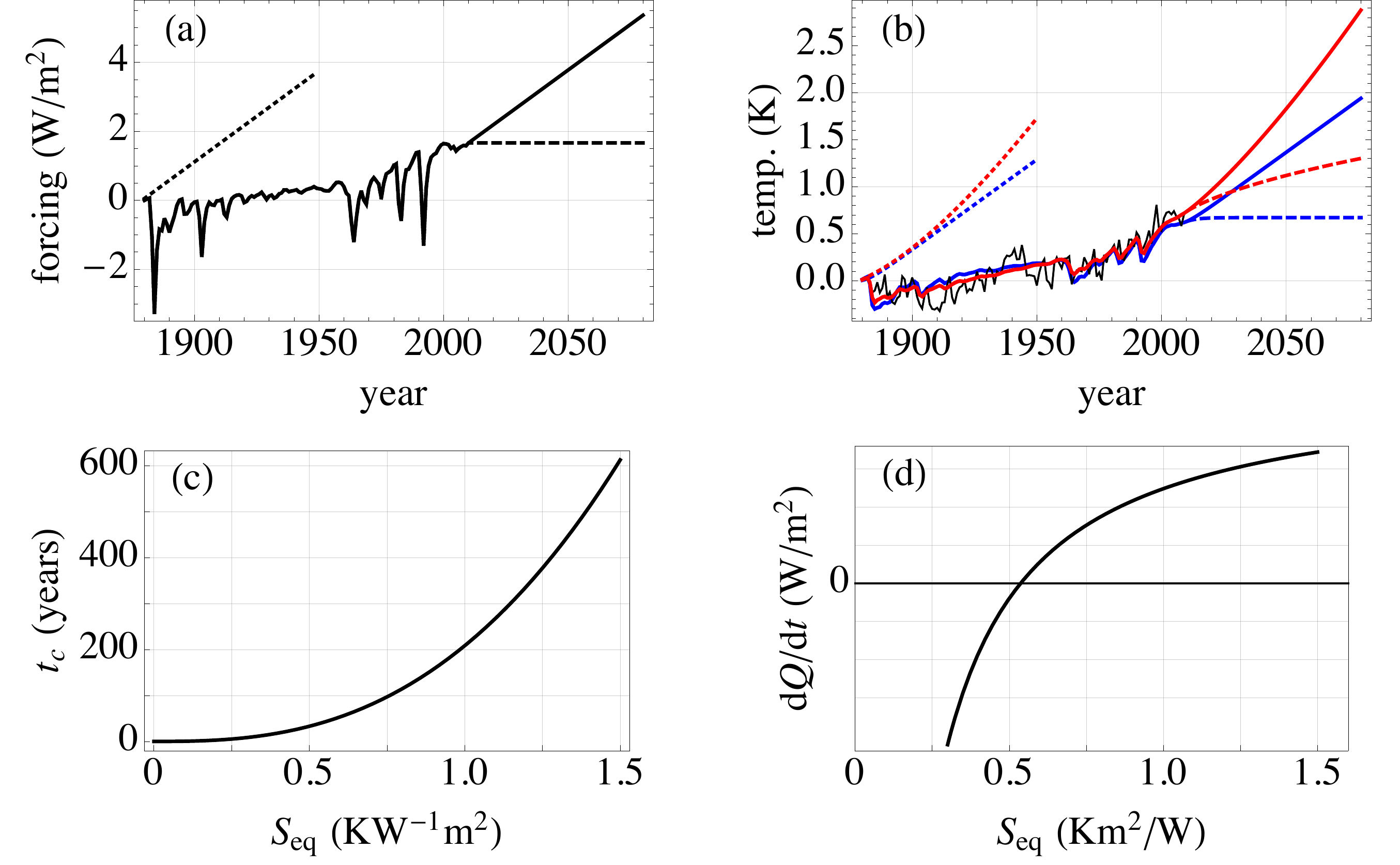}
\caption{(a): Dotted line is the forcing scenario corresponding to 1\% increase in CO$_2$ concentration per year starting in 1880. Solid curve is the historical forcing 1880-2010 followed by 1 \% per year increase in CO$_2$ concentration after 2010. Dashed line is forcing after 2010 kept constant  at the 2010-level. (b): GMST evolution according to the two response models with parameters given in Table 1 for the three forcing scenarios described in (a). Blue curves: exponential response model. Red curves: scale-free response model. (c): The cut-off time $t_c$ versus $S_{\op{eq}}$ as given by equation (\ref{eqA6}) for parameters given in Table 1. (d): $dQ/dt=F-T/S_{\op{eq}}$ at the time of CO$_2$ doubling (year 2080) in the solid-line forcing scenario.}
\end{center}
\end{figure}

As discussed in section~1 this solution and Eq.~(\ref{eqA1}) lead to the rather paradoxical situation of   a negative  energy influx $F-T/S_{\op{eq}}$ resulting from a growing surface temperature for $t-2010>t_c$. We have plotted $t_c$ versus $S_{\op{eq}}$ as determined by Eq.~(\ref{eqA6}) for the estimated values of $\mu$ and  $\beta$ in Fig.~9c. For the median IPCC value $S_{\op{eq}}=0.75$ we have that $t_c\approx 100$ yr, which makes the scale-free model be consistent with a positive energy influx $dQ/dt$ throughout the entire 21st century, even if we assume the validity of Eq.~(\ref{eqA1}). This is also shown in panel (d), where we have plotted the function $F(2080)-T(2080)/S_{\op{eq}}$, i.e., the energy influx versus $S_{\op{eq}}$ at the time of CO$_2$ doubling. From this plot it we observe that the energy influx is positive at this time provided $S_{\op{eq}}>0.5$ Km$^2$/W.

The results shown inFig.~7b, however, suggests that scale-free response is valid at least up to millennium time scale, for which  Eq.~(\ref{eqA1}) and the usual estimates of the magnitude of $S_{\op{eq}}$ would imply simultaneously  negative $dQ/dt$ and continuing rising temperatures after more than a century. Thus, on one hand we have  an empirical model which consistently describes the instrumental  records as well as the reconstruction records for the last millennium, and gives predictions for the 21'st century in agreement with current climate models. On the other hand, predictions made from the model for some imagined future forcing scenarios are inconsistent with conventional notions of the nature and magnitude of equilibrium climate sensitivity. A possible source of this inconsistency can be illustrated by a simple and well-known generalization of the one-box energy-balance model to a two-box model:
\begin{eqnarray}
C_1\frac{dT_1}{dt}&=&-\frac{1}{S_{\op{eq}}}T_1-\kappa ( T_1- T_2)+F\nonumber \\
C_2\frac{d  T_2}{dt}&=&\kappa ( T_1- T_2),  \label{eq14}
\end{eqnarray}
 where $T_1$ could be interpreted as the temperature of the ocean mixed layer, $T_2$ as the temperature of the deep ocean, and $C_1$ and $C_2$ as their respective heat capacities.
 For $C_2\gg C_1$ the 
 response of $T_1$ to a unit step in the forcing is
 \begin{equation}
 R_1(t)\approx S_{\op{tr}}(1-e^{-t/\tau_{\op{tr}}})+(S_{\op{eq}}-S_{\op{tr}})(1-e^{-t/\tau_{\op{eq}}}),\label{eq15}
 \end{equation}
 where $S_{\op{tr}}$, $\tau_{\op{tr}}$ and $\tau_{\op{eq}}$ are given in terms of the parameters of Eq.~(\ref{eq14}) \color{blue} \citep{Rypdal:2012iya}\color{black}.
Two examples of this solution are shown in Fig.~10 for different values of the longer time constant. The solution corresponding to a larger separation between the time constants of the mixed layer and the deep ocean could easily be interpreted incorrectly if integrated only up to $t\sim \tau_{\op{tr}}$, since the  apparent time constant would be $\tau_{\op{tr}}$ and the sensitivity $S_{\op{tr}}$, while the true time constant of the climate system as a whole would be $\tau_{\op{eq}}$ and the true sensitivity would be $S_{\op{eq}}$. This idealized example is of course  only an illustration of the principle that slowly responding components of the climate system and slow feedbacks may obscure the  notion of equilibrium climate sensitivity. The true equilibrium sensitivity (if it exists) may be much larger than estimated from model runs, and hence future warming following a limited period of persistent forcing may be be greater and last longer than predicted from models that do not fully  take into account the LRM-properties  arising from slow responses.

\begin{figure}[h]
\begin{center}
\includegraphics[width=6.5cm]{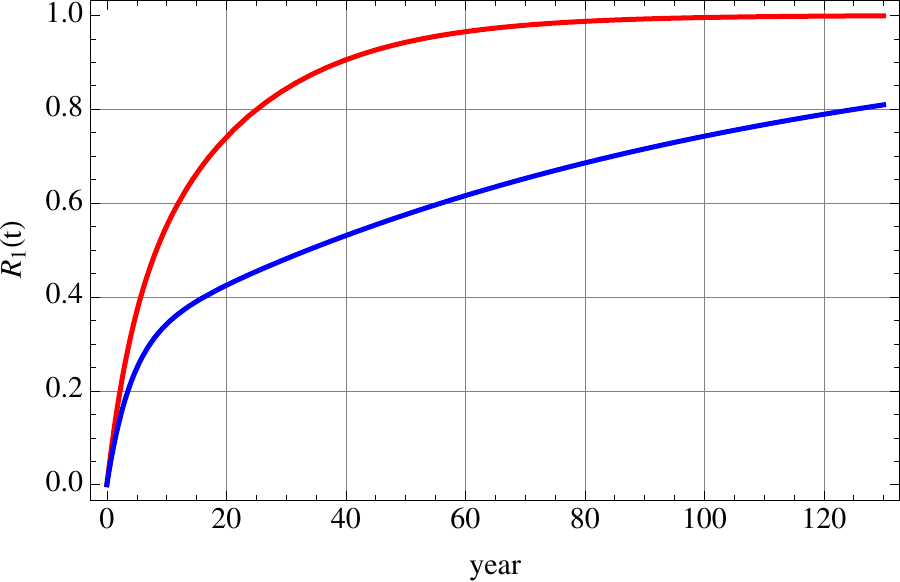}
\caption{The red curve is the response function $R_1(t)$ for the two-box model with $\tau_{\op{tr}}=4$ years, $\tau_{\op{eq}}= 20$ years, $S_{\op{tr}}=0.3$ K/Wm$^{-2}$, and $S_{\op{eq}}=1.0$ K/Wm$^{-2}$. The blue curve  for the same parameters but $\tau_{\op{eq}}= 100$ years.} 
\end{center}
\end{figure}

In a recent paper \color{blue} \cite{Aldrin:2012ht} \color{black} supplemented the information in the  time series of total forcing  and temperatures of the Northern and Southern hemisphere with a series for the evolution of total ocean heat content (OHC) through the last six decades. Their response model is a simple deterministic energy-balance climate/upwelling diffusion ocean model augmented by a first-order autoregressive stochastic process for the residual. The equilibrium climate sensitivity is a parameter in the deterministic model, and since the stochastic term  for the residue is  AR(1) the full model cannot reproduce the LRM properties of the observed climate signal. The purpose of the work is to produce more accurate estimates of $S_{\op{eq}}$, and the introduction of the OHC-data is a new observational constraint on this estimate. We find it interesting to consider these data in the light of a slightly rewritten version of Eq.~(\ref{eqA1});
\begin{equation}
S(t)=\frac{T(t)}{\tilde{F}(t)+F(0)-dQ/dt}, \label{eq16}
\end{equation}
where $S(t)$ can be thought of as a time-dependent climate sensitivity. From the observation data used in  \color{blue} \cite{Aldrin:2012ht} \color{black} we could make  crude linear trend  approximations of OHC, total forcing, and global temperature, i.e.,  $dQ/dt\approx 0.25$ W/m$^{2}$, $\tilde{F}(t)\approx 0.03\, t$ W/m$^{2}$, and $T(t)\approx 0.015\, t$  W/m$^{2}$, where $t$ is time after year 1950  in units of years. Hence we have approximately,
 \begin{equation}
S(t)\approx \frac{1}{2.0+66.7 [F(0)-0.25]/t}. \label{eq17}
\end{equation}

\begin{figure}[h]
\begin{center}
\includegraphics[width=9.0cm]{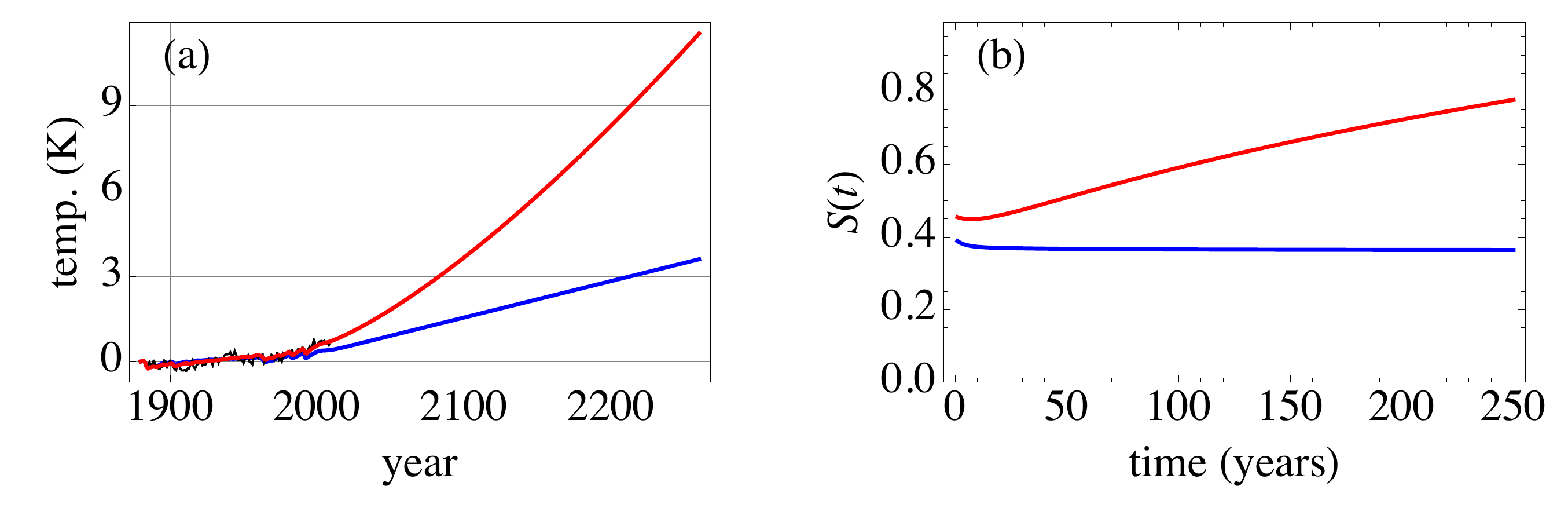}
\caption{(a): Global temperature evolution in response  to the 1\% per yr CO$_2$ increase  forcing scenario, starting in year 2010. Blue curve: the exponential response model. Red curve: the scale-free model. (b): Evolution of the time-dependent climate sensitivity $S(t)$ in response to this forcing scenario assuming a linear increase in OHC corresponding to a net positive energy flux of 0.25 W/m$^2$. The time origin $t=0$ is year 2010. Blue curve: the exponential response model. Red curve: the scale-free model.} 
\end{center}
\end{figure}
In Eq.~(\ref{eq17}) the initial forcing $F(0)$ represents the deviation from radiative equilibrium in 1950, and in order to avoid vanishing sensitivity at $t=0$ we should have $dQ/dt\approx F(0)$ and $S(t)=S_{\op{eq}}\approx 0.5$ Km$^2$/W. Hence these crude trend estimates over the last six decades yield results consistent with the existence of an equilibrium climate sensitivity very close to the best estimate of \color{blue} \cite{Aldrin:2012ht}\color{black}. If we suppose, on the other hand, that the linear trend approximation in temperature is not quite correct, the picture may be different. Consider a linearly increasing  forcing as in the future 1\% CO$_2$ increase scenario shown by the full curve in Figure 9a and the same linear growth $dQ/dt\approx 0.25$ in OHC, but assume that the temperature evolves according to the scale-free response to this forcing shown by the red full curve in Fig.~9b and in Fig.~11a. By inserting these data into Eq.~(\ref{eq16}) we obtain the time-varying climate sensitivity shown in Fig.~11b (here the time origin is chosen in year 2010). Using the temperature evolution for the exponential response shown by the blue curves in Figs.~9b and 11a yields the nearly constant climate sensitivity given by the blue curve in Fig.~11b. This demonstrates that the temperature may increase according to the power-law $\sim t^{\beta/2+1}$ under a linearly increasing forcing and a linearly increasing OHC, provided stronger positive feedback mechanisms take effect on longer time scales and raise the climate sensitivity. In fact, this idea is just a time-domain statement of the concept of a frequency-dependent sensitivity which was formulated in section~3.  The scenario of 1\% increase in CO$_2$ concentration per year  continued 250 years into the future is  a very extreme one, and corresponds to a raise in concentration of more than one order of magnitude. Yet, our results show that within the framework of the scale-free model  the global temperature may increase by more than 10 K while the  OHC maintains a positive linear growth rate throughout this period with only a moderate increase in $S(t)$ from 0.5 to 0.8. The important message from these considerations is that a time- (or frequency-) dependent climate sensitivity may raise the cut-off scale $t_c$ in the scale-free response beyond the millennium time scale, and hence resolves the paradox that the observation data supports the existence of LRM response at least up to such scales.

\section{Discussion and conclusions}
We have in this paper  considered  linear  models of global temperature response  and maximum-likelihood estimation of model parameters. The parameter estimation is  based on observed climate and forcing  records and an assumption of additional stochastic forcing. This modeling shows that a scale-free response is consistent with the stochastic properties of the noisy components of the data, whereas an exponential model is not. The scale-free model with parameters estimated from the modern instrumental  temperature and forcing record successfully predicts the large-scale evolution of the Moberg reconstructed temperature record when the Crowley forcing for the last millennium is used as input. Solutions for the volcanic, solar, and anthropogenic components of the Crowley forcing show that the model ascribes most of the temperature decrease from the MWP to the LIA to the volcanic component, while the rise from  the LIA to year 1979 is attributed to both solar and anthropogenic forcing up till about year 1950 and to anthropogenic after this time. The prediction is not completely independent of the Moberg and Crowley records, however, since the forcing $F(0)$ indicating the departure from climate equilibrium at year 1000, must be estimated from these data. 

%This means that the prediction can be different if a different temperature reconstruction is used. We have made the same analysis on the reconstruction by \color{blue} \cite{Mann:2009kc} \color{black}, but found that for both Moberg and Mann data the estimates of $F(0)$ are very small, indicating that the climate system was very close to equilibrium at the start of the record. Hence the deterministic predictions derived from the two data sets are almost identical. The shift in absolute values of the temperature anomalies between the two reconstructions for the last millennium is due to differences in baseline values for computation of the anomaly. Adjusting for this, the deterministic solution seems to fit the large-scale structure of both reconstructions equally well, which reflects that the two records are mutually quite consistent on the longest scales. The Mann record is stronger low-pass filtered than Moberg making comparison on decadal time scales impossible. On decadal to century time scale the two records share the same LRM characteristics \color{blue}\citep{Rybski:2006bj}\color{black}.

Planet Earth is a non-equilibrium driven physical  system which radiates energy  freely to space. Even when the drive (forcing) is constant in time the total energy content of the system will fluctuate, and if these fluctuations are large on time scales beyond a century it may have little meaning to operate with the notion of an equilibrium climate sensitivity. Thus, the  long-range dependence in the climate response implies  that the equilibrium climate sensitivity concept needs to be generalized to encompass a time-scale dependent sensitivity which incorporates the effect of increasingly delayed positive feedbacks. This may have far-reaching implications for our assessment of future global warming under strong anthropogenic forcing sustained over centuries, as illustrated by the difference between the projected  warming according to the scale-free and  exponential response models shown in Fig.~11a.

The importance of the  ``background'' continuum of time scales in climate variability has been stressed by \color{blue} \cite{Lovejoy:2013vf}\color{black}. In a short review of their own work \color{blue} \cite{Lovejoy:2013te} \color{black}  shows results based on application of their Haar structure function technique to instrumental and multiproxy temperature records.  For instrumental records they find a spectral plateau of $\beta\approx 0.8$ on time scales  up to a decade, but the a sharp transition to $\beta\approx 1.8$ on longer time scales. For the multiproxy records they find a similar transition after a few decades. By similar analysis of ice-core data they also obtain $\beta \approx 1.8$ on time scales greater than a millennium, and argue that this transition constitutes the separation between a macroweather regime to a climate regime. The analysis presented here does not support  that such a transition in the scaling properties of internal variability takes place on decadal time scales in global or hemispheric records. These scaling properties are shown by the wavelet variances of the residuals in Figs. 3b and 7b, and indicate $\beta \approx 0.8$-scaling through the entire instrumental century-long record and the millennium-long multiproxy record, respectively. The transition on multidecadal time scale also fails to show up in the detrending scaling analysis of proxy data in \color{blue} \cite{Rybski:2006bj} \color{black} and \color{blue}\cite{Rypdal:2013cc}\color{black}. We  suggest that the transition reported by \color{blue} \cite{Lovejoy:2013te} \color{black} is a consequence of not distinguishing between forced and stochastic response (alternatively, by not properly eliminating ``trends'' imposed by external forcing). A transition to a more persistent climate regime may perhaps be identified on millienium time scales, but it is an open and interesting question whether the rise of $\beta$ from a stationary ($\beta<1$) to a non-stationary regime ($\beta>1$) is an  actual change in the properties of the climate response or an effect of trends imposed by orbital forcing.

It is important to keep in mind that what we do in this paper is only to select  the optimal   model from two classes of linear response models when the selection criterion is  the model's tendency to reproduce the observed climate record from a given forcing record. It is possible that other classes  would contain models that perform better, and the results are of course not better than the data. The nature of the matter is that the results can be improved from better models, and better and more data, and our understanding is of course incomplete until the delayed feedbacks responsible for the long-range dependence are identified and described from physical principles.

\begin{appendix}
\section*{\begin{center}Maximum-likelihood estimation \end{center}}

In discretized versions the dynamic-stochastic models defined by equation  (\ref{eqA3}) can be written on the form 
$$
{\bf T} = G {\bf F} + {\bf x}\,,
$$
where ${\bf T}=(T_1,T_2,\dots,T_n)$ is the random vector representing the temperature record and ${\bf F}=(F_1,\dots,F_n)$ is the deterministic component of the forcing, i.e. $T_i = T(i \Delta t)$ and $F_i = F(i \Delta t)$ where $\Delta t$ is the time-resolution of the records. The matrix $G$ is defined from the Green's function by $G_{ij} = G((i-j)\Delta t)$, and the vector ${\bf x}=(x_1,\dots,x_n)$ is  a stochastic process. For the exponential response model ${\bf x}$ is an AR(1) process
$$
x_{k+1} = a x_k + \phi\varepsilon_{k},
$$ 
where $\phi=\sigma/C$,  $a=1-1/\tau$,  and $\varepsilon_k$ are i.i.d. Gaussian variables of unit variance. In the scale-free model the process ${\bf x}$ is an fGn. To emphasize the parameter dependence we denote the AR(1) process by ${\bf x}_{C,\sigma,\tau}$ and the fGn by ${\bf x}_{\mu,\sigma, \beta}$. In the same way we denote the Green's function by $G_{C,\tau}$ in the exponential response model, and by $G_{\mu,\beta}$ in the scale-free response model.  

By a simple change of variables the $n$-dimensional probability density function (pdf) of the random vector ${\bf T}$ is related to the pdf of ${\bf x}$ through
$$
p_{\bf T} ({\bf T}) = p_{\bf x}({\bf T}-G{\bf F})\,.
$$ 
 
For temperature observations $\bf T$ the likelihood function for the exponential response model becomes
\begin{equation} \label{L1}
L(C,\sigma,\tau) = p_{\bf x_{C,\sigma,\tau}}({\bf T}-G_{C,\tau}{\bf F})\,,
\end{equation}
and for the scale-free response model: 
\begin{equation} \label{L2}
L(\mu,\sigma,\beta) = p_{\bf x_{\mu,\sigma,\tau}}({\bf T}-G_{C,\tau}{\bf F})\,.
\end{equation}
We see that computation of these likelihoods essentially entails computation of corresponding likelihoods for AR(1) models and fGns. Computation of likelihood functions for auto-regressive processes is straight forward using standard time-series techniques. Effective computation of the likelihood function for fGns can be achieved using the Durbin-Levinson algorithm for inverting the co-variance matrix \color{blue}\citep{McLeod:2007wp}\color{black}.  

In this paper the parameters of the two models are estimated by maximizing Eqs.~ (\ref{L1}) and (\ref{L2}) numerically. 

\end{appendix}

\end{document}